\documentclass[twocolumn,showpacs,superscriptaddress,amsmath,amssymb,aps,pra]{revtex4-1}
\usepackage{amsmath,amsthm,amssymb}
\usepackage{latexsym}
\usepackage{array}
\usepackage{amscd,graphicx}
\usepackage{bm,textcomp}
\usepackage[titletoc]{appendix}
\usepackage{epsfig}
\usepackage{epstopdf}
\usepackage[normalem]{ulem}
\usepackage[dvipsnames]{xcolor}
\usepackage[colorlinks=true,linkcolor=blue,citecolor=blue]{hyperref}

\begin{document}
	
	\newcommand{\red}[1]{{\textcolor{red}{#1}}}
	\newcommand{\qn}[1]{{\textcolor{red}{ (?)  #1 }}}
	\newcommand{\blue}[1]{{\textcolor{blue}{#1}}}
	\newcommand{\cmt}[1]{{\textcolor{blue}{[#1]}}}
	\newcommand{\del}[1]{{\textcolor{green}{ \sout{#1}}}}
	\newcommand{\rvs}[1]{{\textcolor{purple}{#1}}}

\title{Amplification of the coupling strength in a hybrid quantum system}
\author{Wei Xiong}
\affiliation{Quantum Physics and Quantum Information Division, Beijing Computational
	Science Research Center, Beijing 100193, China}
\author{Yueyin Qiu}
\affiliation{Department of Physics, Zhejiang University, Hangzhou 310027, China}
\affiliation{School of Science, Chongqing University of Posts and Telecommunications, Chongqing 400065, China}
\author{Lian-Ao Wu}
\affiliation{Department of Theoretical Physics and History of Science, The Basque Country
University(EHU/UPV), 48080 Bilbao, Spain}
\affiliation{IKERBASQUE, Basque Foundation for Science, 48011 Bilbao, Spain}
\author{J. Q. You}
\altaffiliation[Corresponding author.~jqyou@zju.edu.cn]{}
\affiliation{Department of Physics, Zhejiang University, Hangzhou 310027, China}
\affiliation{Quantum Physics and Quantum Information Division, Beijing Computational
Science Research Center, Beijing 100193, China}

\date{\today }

\begin{abstract}
Realization of strong coupling between two different quantum systems is important for fast transferring quantum information between them, but its implementation is difficult in some hybrid quantum systems.
Here we propose a scheme to enhance the coupling strength between a single
nitrogen-vacancy center and a superconducting circuit via squeezing. The
main recipe of our scheme is to construct a unitary squeezing
transformation by directly tuning the specifically-designed superconducting
circuit. Using the experimentally accessible parameters of the circuit, we
find that the coupling strength can be largely amplified by applying the
squeezing transformations to the system. This provides a new path to enhance
the coupling strengths in hybrid quantum systems.
\end{abstract}

\maketitle

\section{Introduction}

Hybrid quantum systems, with the goal of harnessing the advantages of
different subsystems to better explore new phenomena and potentially bring
about novel quantum technologies (see Ref.~\cite{Xiang-RMP,PNAS-Kurizki} for a review),
can have versatile applications in quantum information. Among various hybrid systems,
the nitrogen-vacancy (NV) center in a diamond
coupled to a superconducting circuit has attracted special attention (see, e.g.,
Refs.~\cite{Kubo-10,Schuster-10,Kubo-11,Marcos-10,Zhu-11,Hoffman-11,Saito-13,Twamley-10,Qiu-14}),
because it has distinct advantages, such as high tunability, long coherence
time, and stable energy levels.  {In addition, superconducting circuits exhibit macroscopic quantum coherence, promise good scalability, and
can be conveniently controlled and manipulated via external fields (see, e.g., Refs.~\cite{You-11,Wendin-17}).
}

However, the coupling strength between a
single NV center and a superconducting circuit is too small to coherently
exchange mutual quantum information~\cite{Marcos-10,Zhu-11, Saito-13,Jin-12}. One solution to overcome this drawback is the use of an ensemble
containing a large number (e.g., $N\sim 10^{12}$) of NV centers, where two
lowest collective excitation states of the ensemble encode a qubit (i.e., a
pseudo-spin). Thus, the coupling strength between the NV-center ensemble and
the superconducting circuit can be effectively enhanced by a factor of $%
\sqrt{N}$~\cite{Raizen-89,Petersson-12,Rose-17}. This makes it possible to reach the
strong-coupling regime of the hybrid system. However, it is difficult for
the ensemble to implement direct single-qubit manipulation and also the
coherence time is greatly shortened due to the inhomogeneous broadening \cite%
{Wesenberg-09,Dobrovitski-08,Julsgaard-13}. Therefore, significantly coupling a single NV
center to a superconducting circuit has been longed for.

Here we propose an experimentally feasible method to effectively amplify the
coupling strength between a single NV center and a superconducting circuit.
The main recipe of our scheme is to prepare the unitary one-mode squeezing
transformations. After applying these squeezing transformations to the
hybrid system, the effective coupling strength can be enhanced by two orders of the magnitude
using the experimentally accessible parameters of the circuit.

The methodology dates back to the amplification of Kerr effect~\cite%
{Bartkowiak-14}, where a rather complicated circuit was exploited. Recently,
a simpler squeezing-transformation circuit has been proposed for the cavity
mode to amplify the coupling in an optomechanical system~\cite{Lu-PRL}, but
the generation of the squeezing terms in the system Hamiltonian requires an
additional driven nonlinear medium. Here we specifically design a superconducting
circuit that enables one to engineer the squeezing transformations by directly tuning the circuit.

The paper is organized as follows. Section~\ref{p2} introduces the Hamiltonian of the proposed hybrid quantum system. In Sec.~\ref{p3}, we design two basic gates by tuning the magnetic flux through the smaller loop of the circuit.
In Sec.~\ref{p4}, we use these two basic gates to construct the squeezing operator and then apply the squeezing transformations to amplify the coupling strength between the single NV center and the superconducting circuit. Finally, we give a brief discussion and conclusion in Sec.~\ref{p5}.

\section{The Hybrid quantum system}\label{p2}

\begin{figure}
\includegraphics[scale=0.4]{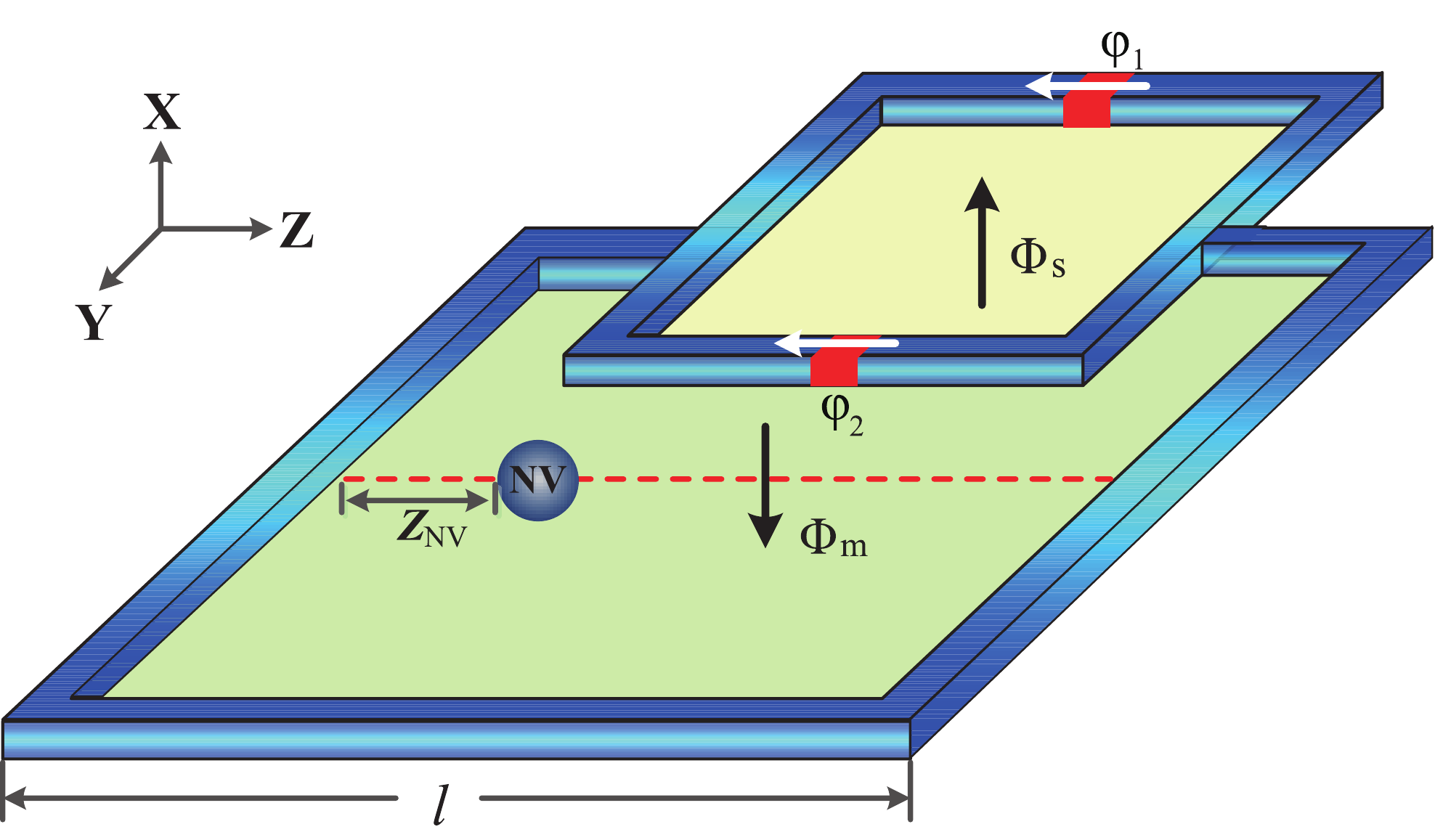}
\caption{(Color online) A superconducting loop  embedding a symmetric SQUID and encircling a
single NV center. For simlicity, we consider a square loop with edge length $l$. The magnetic flux $\Phi_s$ in the SQUID loop is applied in a direction opposite to the magnetic flux $\Phi_m$ in the main
loop. The single NV center is located at $z_{\rm NV}$, which is near
the left edge of the main loop and along the symmetry line (the red dashed
line) in the $z$ direction.}
\label{systemtotal}
\end{figure}

We propose a hybrid system which is composed of a superconducting loop embedding a superconducting quantum interference device (SQUID) and encircling a single NV center (see Fig.~\ref{systemtotal}). Here we consider a symmetric SQUID with identical junction capacitances and Josephson coupling energies, i.e., $C_{1}=C_{2}=C$, and $E_{J1}=E_{J2}=E_{J}$. In addition, we suppose that the main loop of the superconducting circuit is fabricated with a non-negligible inductance $L$, while the SQUID loop is small enough to have a negligible inductance. Also, two static magnetic fields in opposite directions are applied, respectively, to the small and main loops. The fluxoid quantization conditions for these two loops are
\begin{equation}
\varphi_1-\varphi_2+2\pi f_s=0,~~
\varphi_2-2\pi f_m+2\pi IL/\Phi_0=0,  \label{eq1}
\end{equation}
where $f_{s(m)}=\Phi_{s(m)}/\Phi_0$, with $\Phi_0=h/2e$ being the flux quantum, $\varphi_i~(i=1,2)$ is the phase drop across the $i$th Josephson junction in the SQUID, and $I$ is the total circulating current in the main loop.

The kinetic energy of the superconducting circuit corresponds to the
electrostatic energy stored in the capacitors~\cite{Makhlin-01}: $T=\frac{1}{%
	2}C(V_1^{2}+V_2^2)$, where $V_{i}=(\Phi _0/2\pi )\dot{\varphi}_i$
is the voltage across the $i$th Josephson junction in the SQUID. Using the
fluxoid quantization conditions in Eq.~(\ref{eq1}), this kinetic energy can
be written as
\begin{eqnarray}
T\! &\!=\!&\!\frac{1}{2}C\left( \frac{\Phi _{0}}{2\pi }\right) ^{2}\left(
\dot{\varphi}_{1}^{2}+\dot{\varphi}_{2}^{2}\right)  \notag \\
&=&\!C\left( \frac{\Phi _0}{2\pi }\right) ^2\left[ \dot{%
	\varphi}^2+\left( \pi \dot {f}_s\right) ^{2}%
\right],  \label{eq2}
\end{eqnarray}%
where $\varphi \equiv (\varphi_1+\varphi_2)/2$. We consider a static
external flux for $\Phi_s$, so $\dot{f}_{s}=0$. Then, the
kinetic energy $T$ is reduced to $T=C(\Phi _0/2\pi)^2\dot{\varphi}^{2}$.
Also, it follows from Eq.~(\ref{eq1}) that
\begin{equation}
I=-\frac{\Phi_0}{2\pi L}\left(\varphi +\pi f_s-2\pi f_m\right) .  \label{eq3}
\end{equation}%
The inductive energy related to the inductance $L$ is given by
\begin{equation}
U_L\equiv\frac{1}{2}LI^{2}=E_L(\varphi+\pi f_s-2\pi f_m)^2,  \label{eq4}
\end{equation}%
where $E_L=\Phi_0^2/(8\pi^2 L)$. When including this inductive energy, the total potential energy of the superconducting circuit is
\begin{eqnarray}
U\! &\!=\!&\!-E_J(\cos \varphi_1+\cos \varphi_2)+U_L\notag\\
&=&\!-E_{J}(f_s)\cos \varphi +U_L,\label{eq5}
\end{eqnarray}%
where $E_J(f_s)=2E_J\cos (\pi f_s)$ is the flux-dependent effective Josephson energy. The Lagrangian of the superconducting circuit is $\mathcal{L}=T-U$. Assigning $\varphi
$ as the canonical coordinate, we have the canonical momentum $p\equiv\hbar n=\partial \mathcal{L}/\partial \dot{\varphi}=2C\left(\Phi_0/2\pi\right)^2\dot{\varphi}$. Hence the Hamiltonian of the superconducting circuit is given by
\begin{equation}
H_{\rm SC}=E_c n^2-E_{J}(f_s)\cos \varphi+E_L(\varphi+\pi f_s-2\pi f_m)^2,\label{eq6}
\end{equation}
where $E_c=(2e)^2/2C$ is the charging energy of a single Cooper pair and $n=-i\partial/\partial \varphi$ is the number operator of Cooper pairs.

An NV center consists of a substitutional nitrogen atom next to a vacancy in
the diamond lattice \cite{Doherty-13}. It has a spin triplet ground state and a zero-field
splitting $D\approx2.87$ GHz~\cite{Loubser-1978} between the sublevels with the spin $z$ components $m_s=0$
and $m_s=\pm 1$. The strain-induced splitting is
negligible in comparison with the Zeeman effect~\cite{Neumann-09}. In our
proposal, the crystalline axis of the NV center is set as the $z$ direction.
By applying a weak static magnetic field $B_{\mathit{z}}^{\rm ext}$
along the $z$ direction, the two degenerate sublevels $m_s=\pm 1$
are split due to the Zeeman effect. The sublevels $m_s=0$ and $-1$
can be well isolated from other levels by tuning $B_z^{\rm ext}$ and they act as a pseudo-spin. The pseudo-spin Hamiltonian is (we set $\hbar =1$ hereafter)
\begin{equation}
	H_{\rm NV}=\frac{1}{2}\omega_{\rm NV}\tau_z,  \label{eq7}
\end{equation}
where
	\begin{equation}
	\omega_{\rm NV}=D-g_{e}\mu _{B}B_z^{\rm ext}
	\label{eq8}
\end{equation}
is the energy difference between the lowest two sublevels with $m_s=0$ and $-1$, respectively. The corresponding Pauli operators are $\bm{\tau}\equiv (\tau_{x},\tau_{y},\tau_{z})$.
	
As shown in Fig.~\ref{systemtotal}, a single NV center is located at the
coordinate $z_{\rm NV}$, starting from the left edge of the main loop
and along the $z$ direction on the midline. The interaction Hamiltonian
$H_{\rm int}$ of the hybrid system is~\cite{Xiang-13}
\begin{equation}
	H_{\rm int}=-\frac{1}{\sqrt{2}}g_e\mu_B B_x^{\rm SC}(z_{\rm NV})\tau_x,  \label{eq9}
\end{equation}%
where the magnetic field $B_x^{\rm SC}(z_{\rm NV})$ is associated with the persistent current in the main loop. According to the Biot-Savart law, $B_x^{\rm SC}(z_{\rm NV})$ can be written as
\begin{equation}
	B_x^{\rm SC}(z_{\rm NV})=IB_0(z_{\rm NV}),\label{eq10}
\end{equation}%
where
\begin{eqnarray}
		B_0(z_{\rm NV})
		&=&\frac{\mu_0}{4\pi}\Bigg[ \frac{%
		l^2+2 z_{\rm NV}^2}{l z_{\rm NV}\sqrt{(l/2)^{2}+z_{\rm NV}^2}} \notag\\
	    &&+\frac{3l^2-4l z_{\rm NV}+2 z_{\rm NV}^2}{l\left( l-z_{\rm NV}\right) \sqrt{\left( l-z_{\rm NV}\right) ^2+(l/2)^2}} \Bigg].~~~~\label{eq11}
\end{eqnarray}
The total Hamiltonian $H$ of the hybrid quantum system is given by $H=H_{\rm SC}+H_{\rm NV}+H_{\rm int}$.
	
\section{Two basic gates}\label{p3}

We tune the external magnetic field $B_z^{\rm ext}$ to have $%
\omega_{\rm NV}=0$, so as to achieve the two basic gates for
constructing squeezing operations. Denote $\omega_{\rm sc}$ as the
transition frequency between the lowest two energy levels of the
superconducting circuit and $g$ as the coupling strength between the
single NV center and the superconducting circuit. Now the two subsystems
become effectively decoupled due to $|g/(\omega_{\rm sc}-\omega_{\rm NV})|= |g/\omega_{\rm sc}|\ll 1$. Also, we tune the
two external magnetic fields in opposite directions to satisfy $\Phi_m-\Phi_s/2=0 $. Because $\omega_{\rm NV}=0$ and
$|g/\omega _{\rm sc}|\ll 1$, the total Hamiltonian can be
approximately written as
\begin{equation}
H\approx H_{\rm sc}=E_{c}n^{2}-E_{J}(f_s)\cos \varphi
+E_{L}\varphi ^{2}.  \label{eq12}
\end{equation}%
Note that if $L\rightarrow 0$, $E_L\rightarrow\infty$, so it is required that $\varphi \rightarrow 0$ in Eq.~(\ref{eq12}). However, $L\neq 0$ for a realistic circuit. Thus, in this nonzero $L$ case, the phase drop $\varphi $ is not constrained by the loop inductance but mainly by the effective Josephson energy of the SQUID.

By tuning the magnetic flux in the SQUID loop (now denoted as $\Phi_s^{(0)}$) to $\Phi_s^{(0)}=\Phi _{0}/2$, i.e., $f_s^{(0)}\equiv\Phi_s^{(0)}/\Phi_0=1/2$, one has $%
E_J(f_s^{(0)})=0 $, so the Hamiltonian in Eq.~(\ref{eq12}) is
reduced to a harmonic oscillator
\begin{equation}
H_{0}=E_{c}n^{2}+E_{L}\varphi ^{2}.  \label{eq13}
\end{equation}%
In second quantization,
\begin{equation}
\varphi =\sqrt{\frac{1}{2m\omega _{0}}}(a+a^{\dag }),~~n=i\sqrt{\frac{%
		m\omega _{0}}{2}}(a^{\dag }-a),\label{eq14}
\end{equation}%
where $m=1/(2E_c)$, and
\begin{equation}
\omega _{0}=2\sqrt{E_{c}E_{L}}\label{eq15}
\end{equation}%
is the angular frequency of the harmonic oscillator. The creation
(annihilation) operator $a^{\dag }$ ($a$) obeys the bosonic commutation
relation $\left[ a,a^{\dag }\right] =1$, and the Hamiltonian in Eq.~(\ref{eq13}) can be written as
\begin{equation}
H_{0}=\omega _{0}a^{\dag }a.  \label{eq16}
\end{equation}%
Evolving the hybrid system for a time $t$, a quantum gate
\begin{equation}
U_{0}(t)\equiv e^{-iH_{0}t}=e^{-i\omega _{0}a^{\dag }at}  \label{eq17}
\end{equation}%
is achieved.

\begin{figure}
\includegraphics[scale=0.65]{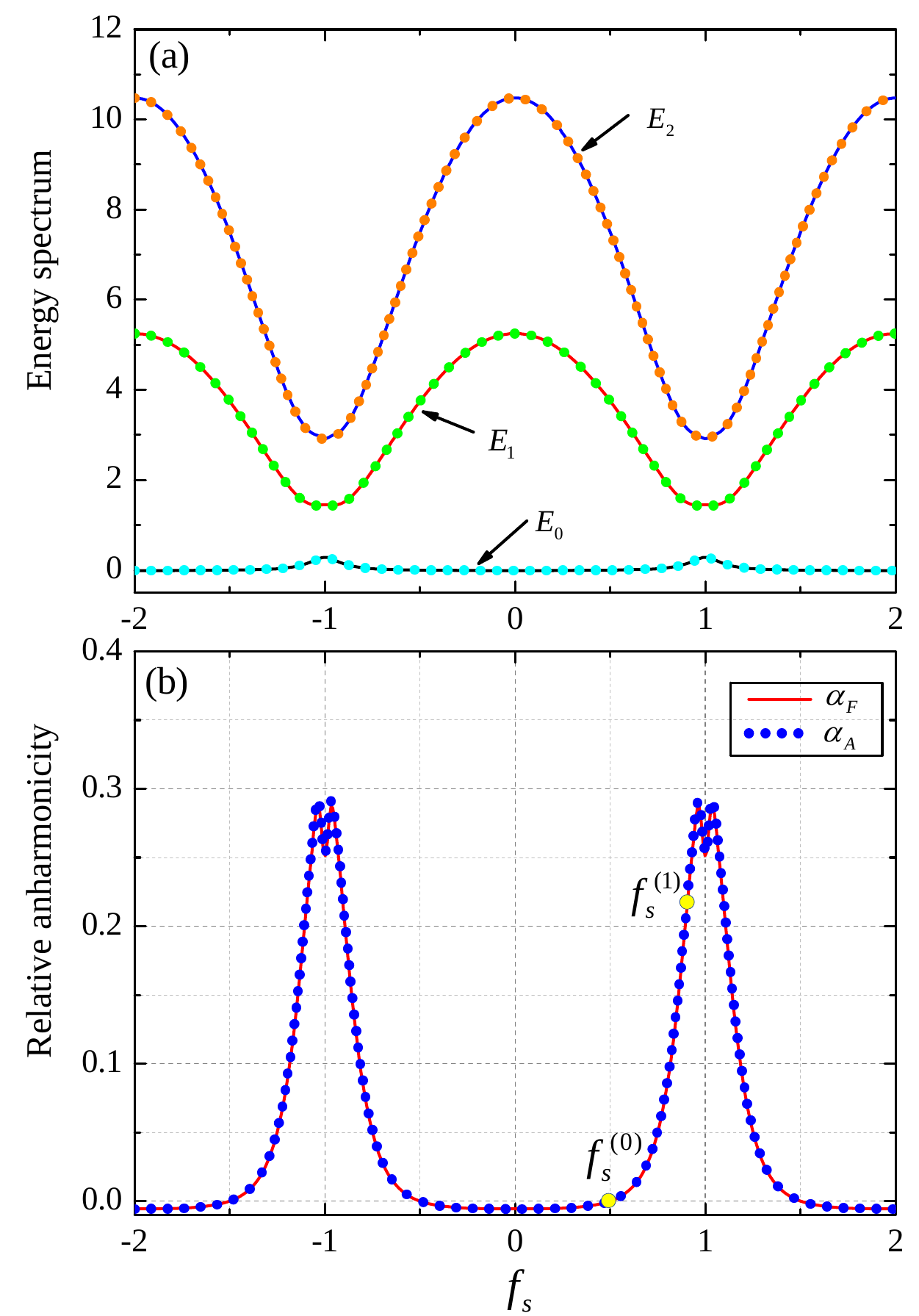}
\caption{(Color online) (a) The lowest three energy levels of the superconducting circuit versus the normalized magnetic flux $f_s$, where $E_0$ is the ground state energy, $E_1$ is the first excited state energy, and $E_2$ is the second excited state energy. The solid and dotted curves are obtained using the Hamiltonians in Eq.~(\ref{eq12}) and Eq.~(\ref{eq19}), respectively.  (b) The relative anharmonicity $\alpha$ versus the normalized magnetic flux $f_s$ when using the Hamiltonians in Eq.~(\ref{eq12}) and Eq.~(\ref{eq19}). The parameters of the circuit are chosen to be $E_{c}=0.12$~GHz, $E_{J}=58$~GHz, and $E_L=58.6$ GHz (corresponding to $L=1.4$~nH).}
\label{nonlinear}
\end{figure}

Here we consider a circuit with $|E_{J}(f_s)/E_{c}|\gg 1$. For this circuit, we can define a quantity $\alpha $ to characterize its
anharmonicity:
\begin{equation}
\alpha =\frac{E_{12}-E_{01}}{E_{01}},\label{eq18}
\end{equation}
where $E_{01}$ is the energy level difference between the ground state energy $E_0$ and
the first excited state energy $E_1$ of the circuit and $E_{12}$ is the energy level
difference between the first and second excited states energies ($E_1$ and $E_2$)  of the circuit. We
use $\alpha _{F}$ to denote the relative anharmonicity of the full
Hamiltonian $H_{\rm sc}$ in Eq.~(\ref{eq12}). Note that the phase $%
\varphi $ is constrained to be small for the circuit with $|E_{J}(f_s)/E_{c}|\gg 1$, so we can write $\cos \varphi \approx 1-\varphi ^2/2!+\varphi ^4/4!$ as a good approximation.
Then, the Hamiltonian $H_{\rm sc}$ in Eq.~(\ref%
{eq12}) is reduced to
\begin{align}
H_1\approx E_c n^2+\frac{1}{2}[2E_L+E_J (f_s)]\varphi^2-\frac{1}{4!}E_J(f_s)\varphi^4.  \label{eq19}
\end{align}
For this approximated Hamiltonian, we use $\alpha_{A}$ to denote its
relative anharmonicity. In Fig.~\ref{nonlinear}(a), we show the lowest three energy levels of the circuit as a function of the normalized magnetic flux $f_s$ in the SQUID loop, where the solid and dotted curves are calculated using the Hamiltonians in Eq.~(\ref{eq12}) and Eq.~(\ref{eq19}), respectively. The parameters are chosen to be $E_{c}=0.12$~GHz, $E_{J}=58$~GHz, and $E_L=58.6$ GHz (corresponding to $L=1.4$~nH~\cite{Sullivan-13}). In Fig.~\ref{nonlinear}(b), we also show the dependence of the relative anharmonicity $\alpha _{F}$ ($\alpha _{A}$) on the normalized
magnetic flux $f_s$. From these results, we can see  that the approximate Hamiltonian in Eq.~(\ref{eq19}) well matches the Hamiltonian in Eq.~(\ref{eq12}).

Away from $\Phi_s^{(0)}=\Phi _{0}$/2, where the gate $U_{0}(t)$
is achieved, we again tune the magnetic flux in the SQUID loop (now denoted
as $\Phi_s^{(1)}$) to, e.g., $\Phi_s^{(1)}\approx
0.9\Phi _{0}$ (i.e., $f_s^{(1)}\equiv\Phi_s^{(1)}/\Phi _{0}=0.9$) to obtain another quantum gate. As shown in Fig.~\ref%
{nonlinear}, this flux is sufficiently away from $\Phi_s^{(0)}$, and the
Hamiltonian (\ref{eq12}) can be well approximated by Eq.~(\ref{eq19}) at the
flux $\Phi_s^{(1)}\approx 0.9\Phi _{0}$. Also, the Hamiltonian (%
\ref{eq19}) has a larger relative anharmonicity at this flux. In second
quantization, the quartic anharmonicity $\varphi ^{4}$ in Eq.~(\ref{eq19})
corresponds to the Duffing terms \cite{Lu-15} $(a+a^{\dag })^{4}$, where the
main contributions arise from the double-photon scattering processes, $%
a^{\dag }aaa$ and $a^{\dag }a^{\dag }a^{\dag }a$. We neglect the high-order
four-photon scattering processes $(a^{\dag })^{4}$ and $a^{4}$, and use a
mean-field approximation \cite{Genes-08} $a^{\dag }a\sim \left\langle
a^{\dag }a\right\rangle =N_a$, where $N_a=\left[ \exp (\omega _{\mathrm{sc}%
}/k_{B}T)-1\right] ^{-1}$ under the thermal equilibrium. At a very low
temperature $T$ (e.g., $\sim 20$~mK ), $\omega _{\rm sc}/k_{B}T\gg 1$
and therefore $N_a\approx 0$. The Hamiltonian (\ref{eq19}) can then be reduced
to
\begin{equation}
H_{1}=\omega_1a^\dag a-\eta_1\left( a^2+a^{\dag 2}\right),\label{eq20}
\end{equation}%
where
\begin{eqnarray}
\omega _{1}&=&\sqrt{2E_{C}[2E_{L}+E_{J}(f_s^{(1)})]}-\eta_{1},\notag \\
\eta_{1}&=&\frac{1}{4}\beta (f_s^{(1)})E_{J}(f_s^{(1)}),\label{eq21}
\end{eqnarray}%
with
\begin{equation}
\beta (f_s^{(1)})=\frac{E_{c}}{2[2E_{L}+E_{J}(f_s^{(1)})]}.\label{eq22}
\end{equation}%
Obviously, the two parameters $\omega_1$ and $\eta_1$ are both controllable by the magnetic flux $\Phi_s$.
Owing to the presence of the inductance $L$, $E_J(f_s)$ can reach the regime of $E_{J}(f_s)<0$ for a harmonic oscillator, where we only ensure $E_L+\frac{1}{2}E_J(f_s)\geq0$. However, the oscillator becomes unstable when $E_L+\frac{1}{2}E_J(f_s)<0$.

With the Hamiltonian in Eq.~(\ref{eq20}), by evolving the hybrid system for a time $t$, another quantum
gate
\begin{equation}
U_{1}(t)\equiv e^{-iH_{1}t}=e^{-i[\omega _{1}a^{\dag }a-\eta
	_{1}(a^{2}+a^{\dag 2})]t}  \label{eq23}
\end{equation}%
is then obtained. Note that a series of quantum gates are used to achieve the coupling amplification between the single NV center and the superconducting circuit (see the next section). To have a high fidelity for each quantum gate, sudden switching between successive gates is needed. In the present case,
one should be able to fast tune the magnetic flux in the SQUID loop. Currently, it is easy to implement such sudden switch as quickly as in just $\sim 1$ ns using conventional techniques (see, e.g., \cite{Silveri-15}). With fast developing quantum technologies, much quicker sudden switch is expected to be implementable.

\section{Amplification of the coupling strength}\label{p4}

\subsection{Squeezing operator}
To enhance the coupling between the single NV center and the superconducting circuit, we need to construct a photon-squeezing operator using the two propagators in Eq.~(\ref{eq17}) and Eq.~(\ref{eq23}). With the annihilation operator $a$ and the creation operator $a^\dag$, we can define three operators
\begin{align}
	\Gamma_1&=\frac{1}{2}(a^2+a^{\dagger 2}),\nonumber\\
	\Gamma_2&=\frac{i}{2}(a^2-a^{\dagger 2}),\label{eq24}\\
	\Gamma_3&=a^\dag a+\frac{1}{2},\nonumber
\end{align}
with commutation relations
\begin{equation}
[\Gamma_1,\Gamma_2]=-2i\Gamma_3,~[\Gamma_2,\Gamma_3]=2i\Gamma_1,~[\Gamma_3,\Gamma_1]=2i\Gamma_2.\label{eq25}
\end{equation}
Therefore, three new operators in Eq.~(\ref{eq24}) can be regarded as the three generators of SU(1,1) group that is non-compact and does not have any finite unitary representation.

Following the method used in Ref.~\cite{Bartkowiak-14}, we can write these operators, in a simple two-dimensional non-Hermitian representation, as
\begin{equation}
	\Gamma_1=i\tau_y,~\Gamma_2=-i\tau_x,~\Gamma_3=\tau_z,\label{eq26}
\end{equation}
where $\bm\tau=(\tau_x,\tau_y,\tau_z)$ are Pauli matrices, and then
\begin{align}
	\exp(-i\gamma_3 \Gamma_3 t)&=\cos(\gamma_3t)-i\Gamma_3\sin(\gamma_3t),\nonumber\\
	\exp(-i\gamma_2 \Gamma_2 t)&=\cosh(\gamma_2t)-i\Gamma_2\sinh(\gamma_2t),\label{eq27}\\
	\exp(-i\gamma_1 \Gamma_1 t)&=\cosh(\gamma_1t)-i\Gamma_1\sinh(\gamma_1t),\nonumber
\end{align}
where $\gamma_1,~\gamma_2$ and $\gamma_3$ are three parameters. Also, the propagators in Eq.~(\ref{eq17}) and Eq.~(\ref{eq23}) can be rewritten, respectively, as
\begin{equation}
U_0(t)=\exp(-i\omega_0\Gamma_3 t),~~U_1(t)=\exp(-i\omega_1\Gamma_3t+i2\eta_1\Gamma_1t).\label{eq28}
\end{equation}
Then, we can employ these two propagators to generate another propagator $U_s(t)=\exp(i2\eta_1\Gamma_1t)$, which can be approximately constructed using
\begin{equation}
U'_s(t)=\lim_{M\rightarrow\infty}[U^\dag_0(t'/M)U_1(t/M)]^M,\label{eq29}
\end{equation}
where $t'=\omega_0 t/\omega_1$, and $M$ is the operation times of the gate $U_0$ ($U_1$).

\begin{figure}
\includegraphics[scale=0.35]{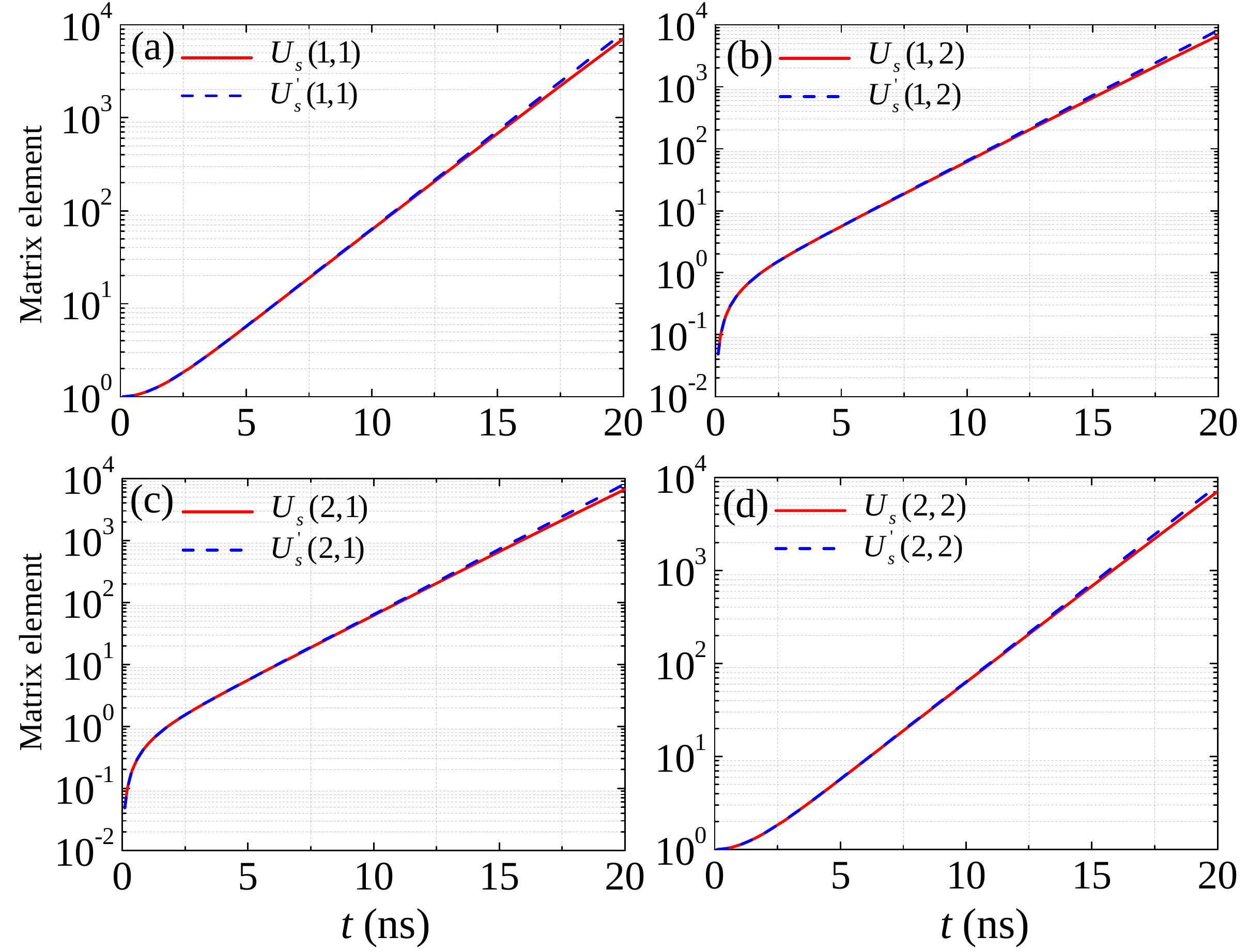}
\caption{(Color online) Numerically obtained matrix elements of $U_s$ and $U_s'$ versus the evolution time $t$, where $M=100$ and $f_s^{(1)}=0.9$. Other parameters are the same as in Fig.~\ref{nonlinear}.}\label{element}	
\end{figure}

Considering the specific parameters used in Fig.~\ref{nonlinear}, we numerically compare the four matrix elements of $U'_s$ with those of $U_s$ and find that $U'_s$ can be well approximated to $U_s$ in the regime of $t/M\leq0.15$ (see Fig.~\ref{element}). Once obtaining the propagator $U_s$, we can combine it and the achieved propagator $U_0$ to produce the desired squeezing operator
\begin{align}
	S&=\exp[\eta_2(a^2-a^{\dagger 2})]\nonumber\\
	&=\exp(-2i\eta_2\Gamma_2)=U_0(t'') U_s(t) U_0^\dagger(t''),\label{eq30}
\end{align}
where $\omega_1t''=(4k+1)\pi/4$, with $k=0,1,2,\ldots$, and $\eta_2=-\eta_1 t$. In the present proposal, we choose the parameters of the circuit to obey $E_J(f_s)<0$, but $E_L+\frac{1}{2}E_J(f_s)>0$, so as to have $\eta_1<0$. To make $\eta_2$ large, we can use a longer evolution time $t$.

\subsection{Coupling enhancement}

Outside the time periods for achieving squeezing transformations $S$ and $%
S^{\dagger }$, we tune $\omega_{\rm NV}$ to be nonzero to satisfy the
near-resonance condition $\omega_{\rm NV}\sim \omega_{\rm sc}$.
In contrast to the sudden switching between successive quantum gates, the
tuning of $\omega_{\rm NV}$ to satisfy the near-resonance condition $%
\omega_{\rm NV}\sim \omega_{\rm sc}$ should be adiabatic. For
the NV center given in Eq.~(\ref{eq7}) and Eq.~(\ref{eq8}), this
adiabatic process can still be achievable by fast tuning the magnetic field
on the NV center, because the level difference has a simple, linear
dependence on the applied magnetic field and no level anticrossing occurs
there.

During this period of near resonance, the coupling between the single
NV center and the superconducting circuit becomes important. Also, the
magnetic flux in the SQUID loop remains at $\Phi_s^{(0)}$.
Therefore, the total Hamiltonian $H_{\rm tot}$ of the hybrid system
reads
\begin{equation}
H_{\mathrm{tot}}=\omega_0 a^\dag a+\frac{1}{2}\omega_{\rm NV}\tau
_{z}+g (a+a^{\dag })\tau _{x},  \label{eq31}
\end{equation}%
where
\begin{equation}
g =\frac{g_{e}\mu _{B}\Phi _{0}{B}_{0}({z}_{\rm NV})[\beta (f_s^{(0)})]^{1/4}}{2\sqrt{2}{\pi L}}.\label{32}
\end{equation}%
To estimate the value of the coupling strength $g$, we choose the
experimentally accessible parameters of the superconducting circuit as in
Fig.~\ref{nonlinear}, i.e., $E_{c}=0.12$~GHz, $E_{J}=58$~GHz, and $L=1.4$~nH.
Using ${z}_{\rm NV}=0.01$~$\mu $m and $f_s^{(0)}=0.5$, we have $g \sim $ $2\pi \times 10$~kHz.

\begin{figure}
	\includegraphics[scale=0.4]{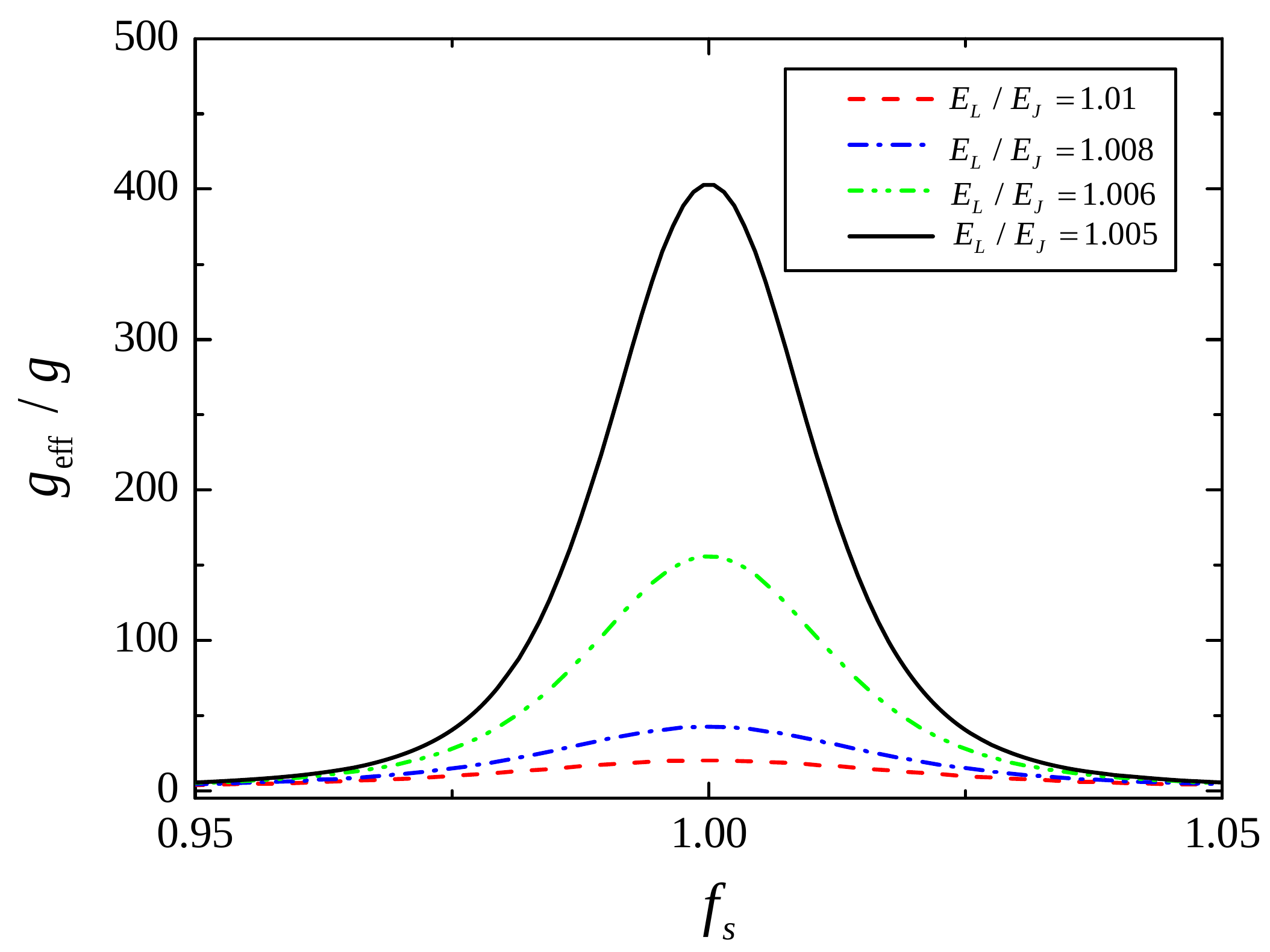}\\
	\caption{(Color online) The effective coupling strength $g_{\rm eff}$ versus the normalized magnetic flux $f_s$ for different ratios of $E_L/E_J$, where $E_c=0.12$ GHz and the time $t$ in Eq.~(\ref{eq34})  is chosen as $t=1$ ns.}\label{amplify}	
\end{figure}

Applying the unitary squeezing transformations $S$ and $S^{\dag }$ to the
Hamiltonian $H_{\rm tot}$, we obtain an effective Hamiltonian for the
hybrid system,
\begin{eqnarray}
H_{\rm eff} &=&SH_{\rm tot}S^{\dag }  \notag \\
&=&\omega_{\rm eff} a^{\dag }a+\frac{1}{2}\omega _{\rm NV}\tau _{z}\notag\\
&&+\chi \left( a^{2}+a^{\dag 2}\right)+g_{\rm eff}(a+a^{\dag })\tau _{x},  \label{eq33}
\end{eqnarray}%
where $\omega_{\rm eff }=\omega_0\cosh \left( 4\eta_2 \right)$ is the transformed frequency of the circuit, $\chi=\frac{1}{2}\omega_0 \sinh(4\eta_2)$ is the strength for squeezing photons, and
\begin{equation}
g_{\rm eff}=g\exp(2\eta_2)\label{eq34}
\end{equation}%
is the effective coupling strength between the single NV center and the superconducting circuit. Obviously, it is enhanced exponentially by a factor of $2\eta_2$.

In Fig.~\ref{amplify}, we plot the effective coupling strength $g_{\rm eff}$ versus the normalized magnetic flux $f_s$ for different ratios of $E_L/E_J$, where $E_c=0.12$ GHz and $t=1$ ns. It shows that the coupling enhancement is very sensitive to the ratio of $E_L/E_J$. Comparing four curves in Fig.~\ref{amplify}, we can see that the coupling strength between the single NV center and the circuit can be enhanced by two orders of the magnitude. Namely, the enhanced coupling strength can approach a few magehertz. Also, it can be further improved by prolonging time or using larger $E_c$, but a long time requires more operation times.

\section{Discussion and Conclusion}\label{p5}

{Without coupling amplification, the coupling strength $g$ between a single NV center and a superconducting circuit can reach $g\approx 2\pi\times 10$~kHz in our proposed hybrid quantum system. This value of the coupling strength was also estimated in Ref.~\cite{Marcos-10}. Experimentally, the coupling strength between a single NV center and the superconducting circuit was reported to be $g\approx 8.8$~kHz in Ref.~\cite{Zhu-11} and $g\approx 4.4$~kHz in Ref.~\cite{Saito-13}. Note that either the theoretically estimated or experimentally achieved value of the coupling strengh is larger than the decoherence rate of a single NV center (which is about $\gamma _{\rm {NV}}\sim1$~kHz in Ref.~\cite{Gaebel-06}), but it is still too weak in comparison with the decoherence rate of the superconducting circuit (which is $\gamma _{\mathrm{sc}}\sim1$~MHz in Ref.~\cite{Zhu-11}). This indicates that the coupling between the single NV center and the superconducting circuit is in the weak-coupling regime, and the decoherence time of this hybrid system is limited by the decoherence time of the superconducting circuit (i.e., $T_{\rm sc}=1/\gamma_{\rm sc}\sim 1$~$\mu$s for $\gamma _{\mathrm{sc}}\sim1$~MHz~\cite{Zhu-11}).}

{For the superconducting circuit system governed by the Hamiltonian (\ref{eq20}), the energy difference of the lowest two levels is about $\Delta E=1.5$~GHz at the point $f_s^{(1)}$ [see Fig.~\ref{nonlinear}(a)]; the corresponding characteristic time of the system is $T_c=1/\Delta E\approx0.67$~ns.  As demonstrated in Ref.~\cite{Gaebel-06}, the typical $\pi/2$- and $\pi$-pulse durations of manipulating an NV center are $15$~ns and $30$~ns, respectively, which are much longer than the characteristic time $T_c$. Thus, manipulating the frequency of an NV center ($\omega_{\rm NV}$) in resonance with the frequency of the superconducting circuit ($\omega_{\rm sc}$) can be nearly adiabatic. Moreover, the typical pulse durations~\cite{Gaebel-06} are much shorter than the decoherence time of the superconducting circuit. Tuning $\omega_{\rm NV}$ to be resonant with $\omega_{\rm sc}$ can be implemented before the system decoheres. In our superconducting circuit, the Josephson coupling energy $E_J$ should be larger than the charging energy $E_c$ and a superconducting loop is introduced. These characteristics are analogous to those of a flux qubit. A recent experiment~\cite{Yan-16} shows that the decoherence time of the flux qubit can be increased to $85$~$\mu$s when shunting a large capacitor to the smaller Josephson junction of the circuit to reduce the effect of the charge noise~\cite{You-07}. This idea can be applied to the superconducting circuit here to improve its quantum coherence.}

{It is shown in Fig.~\ref{amplify} that the effectively coupling strength $g_{\rm eff}$ can be enhanced by two orders of the magnitude when using squeezing transformations. For example, given $g\sim 2\pi \times 10$~kHz, when the parameters in Fig.~\ref{amplify} are used, $g_{\rm eff}$ is enhanced to $\sim 2\pi \times 4$~MHz at $f_s\sim 1$. It has reached a few megahertz.}

In conclusion, we have proposed an experimentally feasible method to
effectively enhance the coupling strength between a single NV center and a
superconducting circuit. The main recipe of our scheme is to use the unitary
squeezing transformations constructed by system evolution. This idea dates back to the amplification of Kerr effect and it can provide a new path to enhance the coupling strengths in hybrid quantum systems.

\section*{ACKNOWLEDGMENTS}

This work is supported by the National Key Research and Development Program of China (Grant No. 2016YFA0301200), the National Natural Science Foundation of China (Grant No.~11774022 and Grant No.~11404019), the National Basic Research Program of China (Grant No.~2014CB921401), the NSAF (Grant No.~U1530401), and the Postdoctoral Science Foundation of
China (Grant No.~2016M600905). L.A.W. is supported by the Basque Government (grant IT472-10) and the Spanish MICINN (Project No.~FIS2012-36673-C03-03). 

W.X. and Y.Q. contributed equally to this work.


\begin{thebibliography}{99}

\bibitem{Xiang-RMP} Z. L. Xiang, S. Ashhab, J. Q. You, and F. Nori, Hybrid quantum circuits: Superconducting circuits interacting with other quantum systems, Rev.
Mod. Phys. \textbf{85}, 623 (2013).

\bibitem{PNAS-Kurizki}G. Kurizki, P. Bertet, Y. Kubo, K. M\o{}lmer, D. Petrosyan, P. Rabl, and J. Schmiedmayer, Quantum technologies with hybrid systems, Proc. Natl. Acad. Sci. USA {\bf 112}, 3866 (2015).

\bibitem{Kubo-10} Y. Kubo, F. R. Ong, P. Bertet, D. Vion, V. Jacques, D.
Zheng, A. Dr\'{e}au, J. F. Roch, A. Auffeves, F. Jelezko, J. Wrachtrup, M.
F. Barthe, P. Bergonzo, and D. Esteve, Strong coupling of a spin ensemble to a superconducting resonator, Phys. Rev. Lett. \textbf{105}, 140502
(2010).

\bibitem{Schuster-10} D. I. Schuster, A. P. Sears, E. Ginossar, L. DiCarlo,
L. Frunzio, J. J. L. Morton, H. Wu, G. A. D. Briggs, B. B. Buckley, D. D.
Awschalom, and R. J. Schoelkopf, High-cooperativity coupling of electron-spin ensembles to superconducting cavities, Phys. Rev. Lett. \textbf{105}, 140501
(2010).

\bibitem{Kubo-11} Y. Kubo, C. Grezes, A. Dewes, T. Umeda, J. Isoya, H.
Sumiya, N. Morishita, H. Abe, S. Onoda, T. Ohshima, V. Jacques, A. Dr\'{e}%
au, J. F. Roch, I. Diniz, A. Auffeves, D. Vion, D. Esteve, and P. Bertet,  Hybrid quantum circuit with a superconducting qubit coupled to a spin ensemble, Phys. Rev. Lett. \textbf{107}, 220501 (2011).

\bibitem{Marcos-10} D. Marcos, M. Wubs, J. M. Taylor, R. Aguado, M. D.
Lukin, and A. S. S\o rensen, Coupling nitrogen-vacancy centers in diamond to superconducting flux qubits, Phys. Rev. Lett. \textbf{105}, 210501 (2010).

\bibitem{Zhu-11} X. Zhu, S. Saito, A. Kemp, K. Kakuyanagi, S. I Karimoto,
H. Nakano, W. J. Munro, Y. Tokura, M. S. Everitt, K. Nemoto, M. Kasu, N.
Mizuochi, and K. Semba, Coherent coupling of a superconducting flux qubit to an electron spin ensemble in diamond, Nature (London) \textbf{478}, 221 (2011).

\bibitem{Hoffman-11} J. E. Hoffman, J. A. Grover, Z. Kim, A. K. Wood, J. R.
Anderson, A. J. Dragt, M. Hafezi, C. J. Lobb, L. A. Orozco, S. L. Rolston,
J. M. Taylor, C. P. Vlahacos, and F. C. Wellstood, Atoms talking to SQUIDs, Rev. Mex. F\'{\i}s. S
\textbf{57}, 1 (2011).

\bibitem{Saito-13} S. Saito, X. Zhu, R. Ams\"{u}ss, Y. Matsuzaki, K.
Kakuyanagi, T. Shimo-Oka, N. Mizuochi, K. Nemoto, W. J. Munro, and K. Semba, Towards realizing a quantum memory for a superconducting qubit: Storage and retrieval of quantum states,
Phys. Rev. Lett. \textbf{111}, 107008 (2013).

\bibitem{Twamley-10} J. Twamley and S. D. Barrett, Superconducting cavity bus for single nitrogen-vacancy defect centers in diamond, Phys. Rev. B \textbf{81},
241202 (2010).

\bibitem{Qiu-14} Y. Qiu, W. Xiong, L. Tian, and J. Q. You, Coupling spin ensembles via superconducting flux qubits, Phys. Rev. A
\textbf{89}, 042321 (2014).

\bibitem{You-11}J. Q. You and F. Nori, Atomic physics and quantum optics using superconducting circuits, Nature \textbf{474}, 589 (2011).

\bibitem{Wendin-17} G. Wendin, Quantum information processing with superconducting circuits: a review,
Rep. Prog. Phys. \textbf{80}, 106001 (2017).

\bibitem{Jin-12} P. Q. Jin, M. Marthaler, A. Shnirman, and G. Sch\"{o}n, Strong coupling of spin qubits to a transmission line resonator, Phys. Rev. Lett. \textbf{108}, 190506 (2012).

\bibitem{Raizen-89} M. G. Raizen, R. J. Thompson, R. J. Brecha, H. J.
Kimble, H. J. Carmichael, Normal-mode splitting and linewidth averaging for two-state atoms in an optical cavity, Phys. Rev. Lett. \textbf{63}, 240 (1989).

\bibitem{Petersson-12} K. D. Petersson, L. W. McFaul, M. D. Schroer, M.
Jung, J. M. Taylor, A. A. Houck, and J. R Petta, Circuit quantum electrodynamics with a spin qubit, Nature (London) \textbf{490}, 380 (2012).

\bibitem{Rose-17}B. C. Rose, A. M. Tyryshkin, H. Riemann, N. V. Abrosimov, P. Becker, H.-J. Pohl, M. L. W. Thewalt, K. M. Itoh, and S. A. Lyon, Coherent Rabi dynamics of a superradiant spin ensemble in a microwave cavity, Phys. Rev. X {\textbf 7}, 031002 (2017).

\bibitem{Dobrovitski-08} V. V. Dobrovitski, A. E. Feiguin, D. D. Awschalom,
and R. Hanson, Decoherence dynamics of a single spin versus spin ensemble, Phys. Rev. B \textbf{77}, 245212 (2008).

\bibitem{Wesenberg-09} J. H. Wesenberg, A. Ardavan, G. A. D. Briggs, J. J.
L. Morton, R. J. Schoelkopf, D. I. Schuster, and K. M\o lmer, Quantum computing with an electron spin ensemble, Phys. Rev. Lett. \textbf{103}, 070502 (2009).

\bibitem{Julsgaard-13}B. Julsgaard, C. Grezes, P. Bertet, and K. M\o lmer, Quantum memory for microwave photons in an inhomogeneously broadened spin ensemble, Phys. Rev. Lett. \textbf {110}, 250503 (2013).

\bibitem{Bartkowiak-14} M. Bartkowiak, L. A. Wu, A. Miranowicz, Quantum circuits for amplification of Kerr nonlinearity via quadrature squeezing, J. Phys. B:
At. Mol. Opt. Phys. \textbf{47}, 145501 (2014).

\bibitem{Lu-PRL} X. Y. L\"{u}, Y. Wu, J. R. Johansson, H. Jing, J. Zhang,
and F. Nori, Squeezed optomechanics with phase-matched amplification and dissipation, Phys. Rev. Lett \textbf{114}, 093602 (2015).

\bibitem{Makhlin-01} Y. Makhlin, G. Sch{\"{o}}n, and A. Shnirman, Quantum-state engineering with Josephson-junction devices, Rev. Mod.
Phys. \textbf{73}, 357 (2001).

\bibitem{Doherty-13} M. W. Doherty, N. B. Manson, P. Delaney, F. Jelezko, J. Wrachtrup, L. C. L. Hollenberg, The nitrogen-vacancy colour centre in diamond, Phys. Rep. \textbf{528}, 1(2013).

\bibitem{Loubser-1978} J. H. N. Loubser, J. A. van Wyk, Electron spin resonance in the study of diamond, Rep. Progr. Phys. \textbf{41}, 1201 (1978).

\bibitem{Neumann-09} P. Neumann, R. Kolesov, V. Jacques, J. Beck, J. Tisler,
A. Batalov, L. Rogers, N. B. Manson, G. Balasubramanian, F. Jelezko, and J.
Wrachtrup, Excited-state spectroscopy of single NV defects in diamond using optically detected magnetic resonance, New. J. Phys. \textbf{11}, 013017 (2009).

\bibitem{Xiang-13} Z. L. Xiang, X. Y. L\"{u}, T. F. Li, J. Q. You, and F.
Nori, Hybrid quantum circuit consisting of a superconducting flux qubit coupled to a spin ensemble and a transmission-line resonator, Phys. Rev. B \textbf{87}, 144516 (2013).

\bibitem{Sullivan-13} D. F. Sullivan, S. K. Dutta, M. Dreyer, M. A. Gubrud,
A. Roychowdhury, J. R. Anderson, C. J. Lobb, and F. C. Wellstood, Asymmetric superconducting quantum interference devices for suppression of phase diffusion in small Josephson junctions, J. Appl.
Phys. \textbf{113}, 183905 (2013).

\bibitem{Lu-15} X. Y. L\"{u}, J. Q. Liao, L. Tian, and F. Nori, Steady-state mechanical squeezing in an optomechanical system via Duffing nonlinearity, Phys. Rev. A \textbf{91}, 013834 (2015).


\bibitem{Genes-08} C. Genes, D. Vitali, P. Tombesi, S. Gigan, and M.Aspelmeyer, Ground-state cooling of a micromechanical oscillator: Comparing cold damping and cavity-assisted cooling schemes, Phys. Rev. A \textbf{77}, 033804 (2008).


\bibitem{Silveri-15} M. P. Silveri, K. S. Kumar, J. Tuorila, J. Li, A. Veps \"{a}l\"{a}inen, E. V. Thuneberg and G. S. Paraoanu, St\"{u}ckelberg interference in a superconducting qubit under periodic latching modulation, New. J. Phys. \textbf{17}, 043058 (2015).

\bibitem{Gaebel-06} T. Gaebel, M. Domhan, I. Popa, C. Wittmann, P. Neumann,
F. Jelezko, J. R. Rabeau, N. Stavrias, A. D. Greentree, S. Prawer, J.
Meijer, J. Twamley, P. R. Hemmer, and J. Wrachtrup, Room-temperature coherent coupling of single spins in diamond, Nat. Phys. \textbf{2}, 408 (2006).

\bibitem{Yan-16}
F. Yan, S. Gustavsson, A. Kamal, J. Birenbaum, A. P. Sears, D. Hover, T. J. Gudmundsen, D. Rosenberg, G. Samach,
S. Weber, J. L. Yoder, T. P. Orlando, J. Clarke, A. J. Kerman, and W. D. Oliver,
The flux qubit revisited to enhance coherence and reproducibility,
Nat. Commun. \textbf{7}, 12964 (2016).

\bibitem{You-07}
J. Q. You, X. Hu, S. Ashhab, and F. Nori, Low-decoherence flux qubit, Phys. Rev. B \textbf{75}, 140515 (2007).


\end{thebibliography}
\end{document}